\documentclass[doublecol]{epl2}
\title{Rational material design of mixed-valent high T$_c$ superconductors} 
\author{Z. P. Yin\inst{1} \and G. Kotliar\inst{1}}
\shortauthor{Z. P. Yin and G. Kotliar} %\etal}

\institute{
  \inst{1} Department of Physics and Astronomy, Rutgers University, Piscataway, New Jersey 08854, United States.\\
}

\abstract{
We design, from first principles calculations, a novel family of thallium halide-based
compounds as candidates for new high temperature superconductors, whose superconductivity
is mediated by the recently proposed mechanism of non-local correlation-enhanced
strong electron-phonon coupling. Two prototype compounds namely CsTlF$_3$ and CsTlCl$_3$
are studied with various hole doping levels and volumes. The critical superconducting temperature
T$_c$ are predicted to be about 30 K and 20 K with $\sim$0.35/f.u. hole doping and require
only modest pressures ($\sim$10 and $\sim$2 GPa), respectively.
Our procedure of designing this class of superconductors is quite
general and can be used to search for other ``other high
temperature superconductors".
}

\begin{document}

\maketitle

\section{Introduction}
The holy grail of theory and computation
assisted material design is to accelerate the discovery and
synthesis of new materials with specific desirable properties.
With the rapid development of theory, algorithms, computer
codes, and high-performance computers in the recent years, this
goal is within reach for materials and properties which are firmly
understood with well established tools. In the past two decades,
first-principles calculations have been used to look for better
functional materials such as batteries and energy storage
materials\cite{Ceder, Menga}, spintronics\cite{Sato}, alloy
catalysts\cite{Greeley}, multiferroic and magnetoelectric
materials\cite{Stengel, fennie}, etc. The topological insulators
and Weyl semimetals, which are under current extensive
investigations, were first predicted by theory and
first-principles calculations.\cite{TI1, TI2, Weyl}

An even more ambitious goal is to approach computational
material design in problems, such as reaching higher temperature
superconductivity, where there is yet no consensus on the basic
mechanisms operating in various classes of materials. Still, given an
understanding of a mechanism, modern electronic structure tools
can be put to use, and this process in conjunction with subsequent
experiments, serves to test our basic understanding of the
phenomena and the state of the art of existing algorithms and codes.
Numerous attempts have been made to look for high T$_c$ superconductors in hydrogen-rich materials
such as SiH$_4$\cite{Feng}, SiH$_4$(H$_2$)$_2$\cite{YLi},
and pure hydrogen\cite{McMahon} under high pressure
as well as MgB$_2$-like materials such as LiBC\cite{Rosner}, to name a few.

In a recent paper\cite{BBO}, Yin \textit{et al.} found that the
electron-phonon coupling in a broad class of materials is
significantly enhanced over the value calculated
by density functional theory (DFT)
in the local density approximation (LDA). This mechanism
explains the presence of superconductivity in
materials whose mechanism is not magnetic, but nevertheless the
standard implementations of the Migdal-Eliashberg theory using
LDA computed parameters fail to account for the high T$_c$s. These
``other high temperature superconductors''\cite{pickett2}, such
as the bismuthates and the transition metal chloronitrides,
superconduct at high temperatures as a result of a
non-local correlation-enhanced strong electron-phonon coupling.\cite{BBO}
The approach proposed in ref.~\cite{BBO} passed several stringent
tests. It is able to explain both high temperature
superconductivity as in Ba$_{1-x}$K$_x$BiO$_3$\cite{Cava} 
and low/vanishing temperature
superconductivity as in the Ba$_{n+1}$Bi$_n$O$_{2n+1}$ family
with small $n$\cite{BBO327}. The variation with ions and doping
dependence of T$_c$ in the HfNCl family\cite{MNX} also has a
natural explanation within the approach.\cite{BBO} This class of
materials superconduct in close proximity to a metal-insulator
transition-a region which is empirically propitious
for high temperature superconductivity. In this region, LDA
overestimates the screening yielding an estimate of the coupling
to the phonon vibrations much smaller than its actual value. An
accurate description of the electronic structures and lattice
dynamics of this class of materials requires treatment of
correlation effects beyond LDA and is fulfilled by the GW method
and the HSE06\cite{HSE2006} \textit{screened} hybrid functional
DFT. To evaluate the realistic electron-phonon coupling, Yin
\textit{et al.}\cite{BBO} proposed a simple but efficient
approach that combines the LDA linear response calculations and
the GW and/or \textit{screened} hybrid functional frozen phonon
calculations.

Motivated by the above study\cite{BBO}, in this letter,
we put this approach to the test, by using it to
search for potential candidates of high temperature superconductors in the mixed-valent class similar to the bismuthates.
We identify a new class of thallium halide compounds which can readily form, and has potential for
superconductivity of about 30 K.

\begin{figure}[htb]
\includegraphics[width=0.80\linewidth]{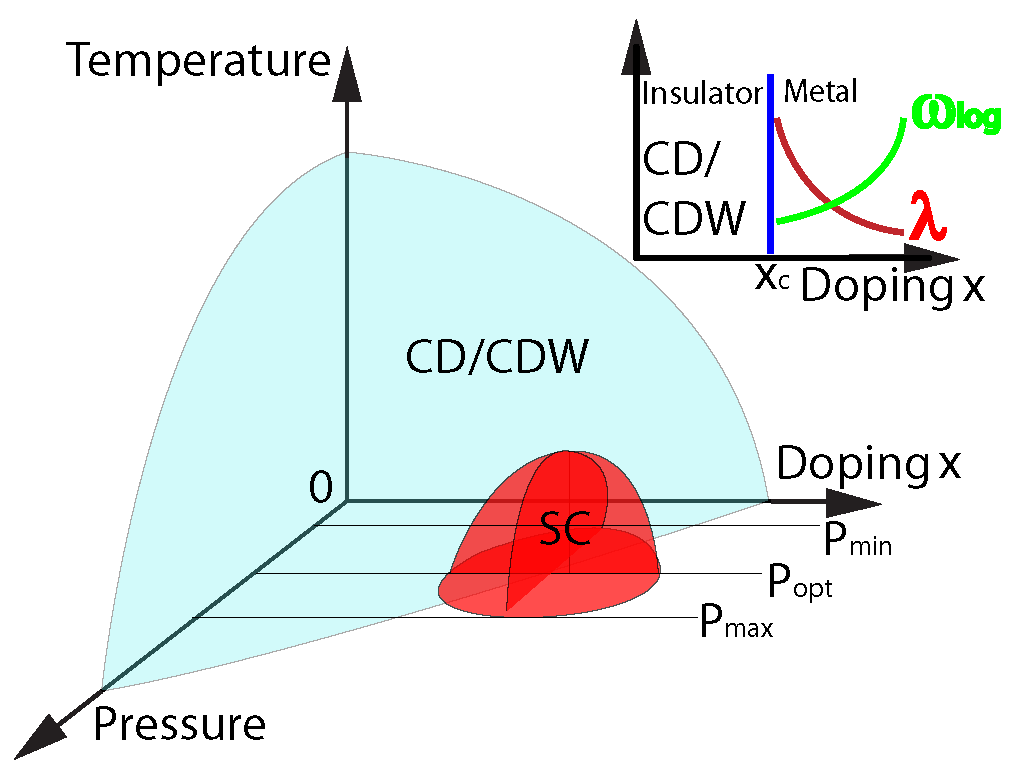}
\caption{
(Color online)
%\textbf{Schematic phase diagram.}
\textbf{The schematic phase diagram of mixed-valence compounds such as BaBiO$_3$ and CsAuCl$_3$.}
At low pressure and small doping, the compounds have structural distortions and are in the
charge disproportionation (CD)/charge density wave (CDW) insulating phase.
With enough doping and pressure,
the structural distortions and charge disproportionation
can be suppressed and they may enter the superconducting (SC) phase.
The insert shows the doping dependence of the electron-phonon coupling $\lambda$
and average phonon frequency $\omega_{log}$ in the metallic phase.
}
\label{phase-diagram}
\end{figure}

\section{General phase diagram}
We search for superconductivity near a charge disproportionation/charge density wave/valence instability.
The generic schematic phase diagram is shown in fig.\ref{phase-diagram} and describes (Ba,K)BiO$_3$.
With small doping and at ambient pressure and low pressure,
the compounds exhibit structure distortions and are in the
charge disproportionation (CD)/charge-density wave (CDW) insulating phase.
With sufficient doping and/or at high pressure, the structure distortions
and charge disproportionation can be suppressed and the compounds enter the metallic phase and
possibly the superconducting phase at low temperature.
Moreover, superconductivity may occur only in some range of doping and pressure.
Therefore, to achieve the highest superconducting T$_c$, both the pressure and doping need to be optimized.

\section{General procedure and computational method}
In this work we look specifically for a class of BaBiO$_3$-like compounds satisfying the following criteria:
(1) the parent compound is an insulator and contains mixed-valent cations 
such as Bi (Bi$^{3+}$ and Bi$^{5+}$), Au (Au$^{1+}$ and Au$^{3+}$) and Tl (Tl$^{1+}$ and Tl$^{3+}$);
(2) the properly doped compound is metallic and has strongly phonon-coupled bands across the Fermi level.
(1) and (2) suggest the compound is a potential ``other high-temperature superconductor'' like (Ba,K)BiO$_3$,
and can be checked by band structure
and linear response calculations where the T$_c$ can be estimated using the approach invoked in ref.~\cite{BBO}.
In addition, we try to determine the energetics of the compounds.
Ideally, one needs to verify that the proposed compound is energetically favorable among all possible competing
phases including elements, binaries, etc.\cite{Zunger}.
However, it is impossible to try all the possibilities.
In practice, we check the following two conditions: a) the proposed compound
should not be very energetically unfavorable relative to the reactants\cite{Zunger}
and b) the proposed compound is dynamically stable, i.e., no unstable phonon modes.

With these considerations in mind, we use density functional theory calculations and linear response calculations
to design new materials, check their stabilities, and estimate the electron-phonon coupling and
the superconducting T$_c$ using the approach proposed in ref.\cite{BBO}.

\begin{figure}[htb]
\includegraphics[width=0.80\linewidth]{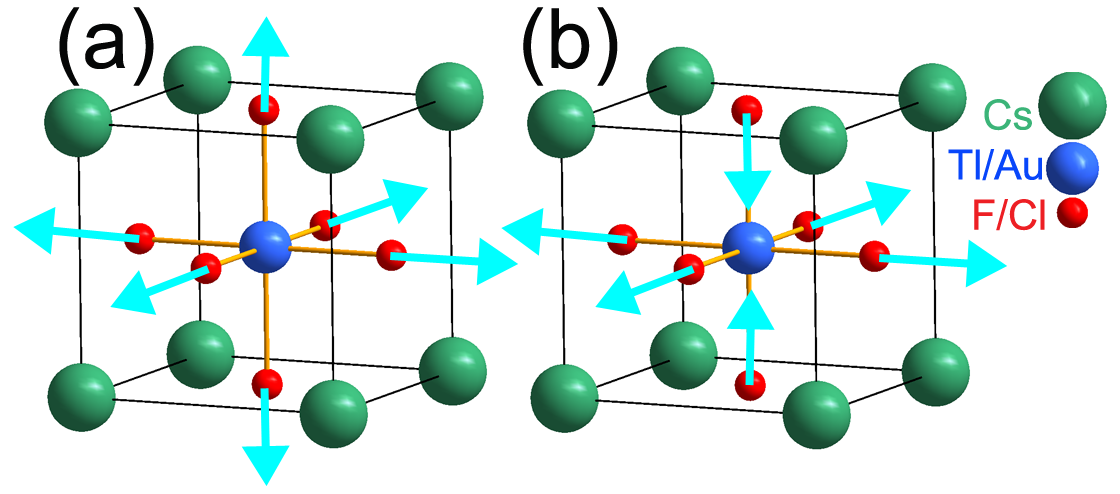}
\includegraphics[width=0.80\linewidth]{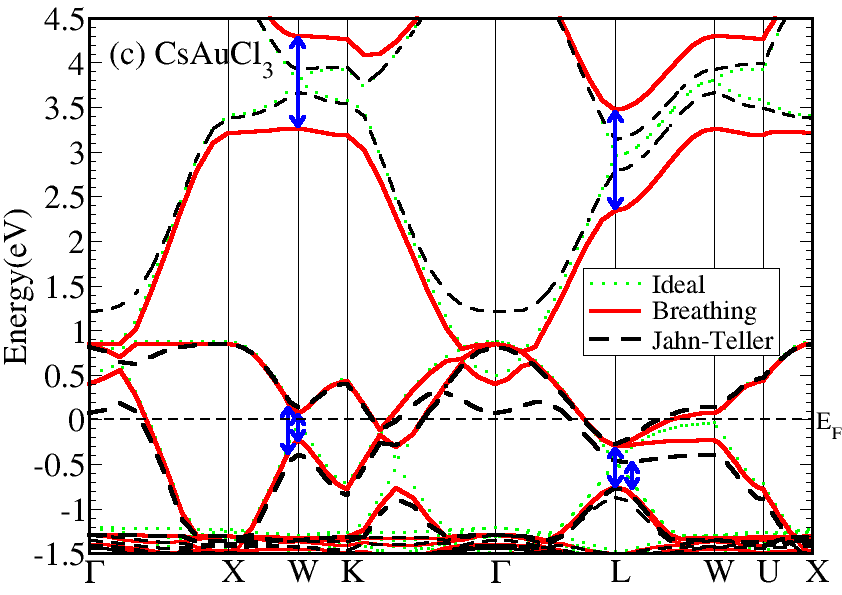}
\includegraphics[width=0.80\linewidth]{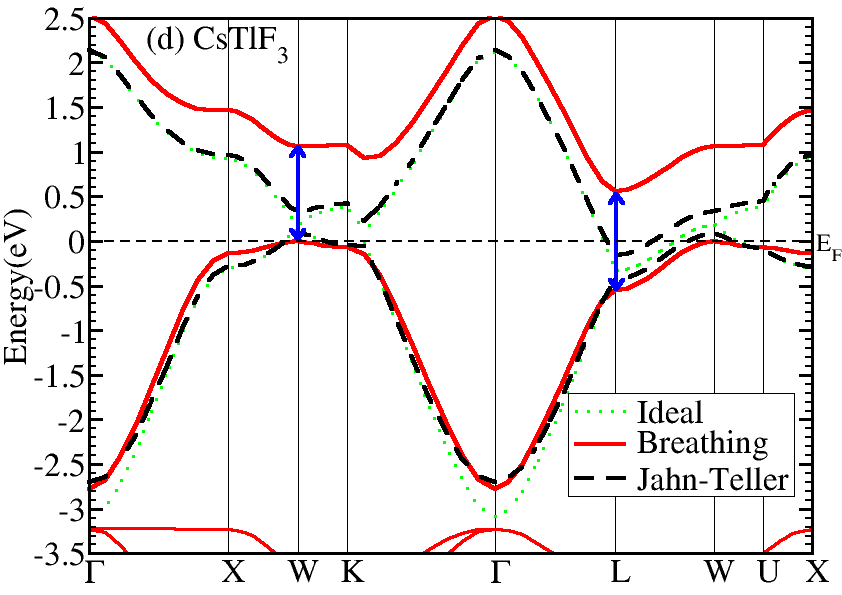}
\caption{
(Color online)
\textbf{The crystal structure and band structures of CsAuCl$_3$ and CsTlF$_3$.}
(a) and (b) shows the crystal structure of simple cubic perovskite CsAuCl$_3$ and CsTlF$_3$ with arrows showing
(a) the F/Cl breathing distortion and (b) the F/Cl Jahn Teller distortion.
The corresponding band structures at ideal structure (simple cubic) and with the F/Cl breathing and Jahn-Teller distortions
are calculated by DFT-GGA and shown in (c) for CsAuCl$_3$ at $a$=5.35$\AA$ and (d) for CsTlF$_3$ at $a$=4.6 $\AA$, respectively.
The arrows in (c) and (d) indicate the magnitudes of the band splittings under the F/Cl breathing and Jahn-Teller distortions,
with atomic displacements of about 0.11 $\AA$ for CsAuCl$_3$ and about 0.09 $\AA$ for CsTlF$_3$ for both distortions.
}
\label{band}
\end{figure}

\section{From starting materials to new materials}
A heuristic reasoning leading to our new materials is the following: 
CsTlCl$_3$ and CsTlF$_3$ are isostructural with BaBiO$_3$ and share
the same valence electron counting as BaBiO$_3$, thus they retain essentially the same band structure around the Fermi level as BaBiO$_3$.
With these critical similarities with BaBiO$_3$, superconductivity is expected when CsTlCl$_3$ and CsTlF$_3$
are optimally doped to suppress the structural distortion.
Considering the isotope effect, CsTlF$_3$ can have better superconductivity than CsTlCl$_3$.

Alternatively, we can start from the existing CsAuCl$_3$ family\cite{CAC} and compare it to the celebrated BaBiO$_3$ family.
While both families crystallize in the (distorted) perovskite structure 
and contain mixed-valent cations Bi (Bi$^{3+}$ and Bi$^{5+}$) and Au (Au$^{1+}$ and Au$^{3+}$),
the parent BaBiO$_3$ compound has oxygen-breathing distortion (arrows in fig.\ref{band}a) whereas the parent CsAuCl$_3$ displays
Cl Jahn-Teller distortion (arrows in fig.~\ref{band}b).
Moreover, unlike the BaBiO$_3$ family with superconducting T$_c$ up to 32 K in optimal doped (Ba,K)BiO$_3$\cite{Cava},
no superconductivity has been found in the CsAuCl$_3$ family
after extensive experimental searches, even under high pressures.\cite{CAC, Fisher}

Since a non-local correlation-enhanced electron-phonon coupling is responsible for the superconductivity in (Ba,K)BiO$_3$,
an evaluation of the electron-phonon matrix elements
of CsAuCl$_3$ is helpful to understand the absence of superconductivity in the CsAuCl$_3$ family.
Figure ~\ref{band}c shows the band structures of CsAuCl$_3$ in the perfect perovskite structure and with
the Cl breathing and Jahn-Teller distortions calculated by VASP\cite{vasp} with
generalized gradient approximation\cite{PBE} (GGA).
With the Cl breathing/Jahn-Teller distortion, the largest band splitting near the Fermi level occurs around $L$/$W$ point, giving rise to
a reduced electron-phonon matrix element (REPME, defined as the ratio of the band splitting at a high symmetry point
to twice the atomic displacement of the corresponding phonon mode, see ref.\cite{BBO} and fig.\ref{band}c) of 2.1/2.5 eV/$\AA$.
These REPMEs \textit{near the Fermi level} in CsAuCl$_3$ are only one third of the DFT-GGA REPME (7.6 eV/$\AA$) of the oxygen-breathing mode in BaBiO$_3$.\cite{BBO}
Such small REPMEs in CsAuCl$_3$ imply weak electron-phonon coupling
thus can hardly host decent superconductivity in the CsAuCl$_3$ family.

However, a good superconductor may still be found in the CsAuCl$_3$-like compounds.
Notice that for the Cl-breathing distortion, the band splittings near $L$ and $W$ points at about 3-4 eV above the Fermi level
are much larger (~5.3 eV/$\AA$), about two and half times the corresponding values near the Fermi level, resulting in
REPMEs comparable to those found in BaBiO$_3$. It suggests a strong coupling of this band (centered at 3 eV above Fermi level)
to the Cl-breathing phonons which can gives rise to a strong total electron-phonon coupling and mediate good superconductivity.
The key is to make this band operate at the phonon energy scale, which
requires the Fermi level of CsAuCl$_3$ to move up by $\sim$3 eV through electron doping ($\sim$2 electrons per formula unit (f.u.)).
Here we search for compounds with the same perovskite structure but with higher Fermi level.
A good choice is to replace Au with Tl
since Tl has two more electrons than Au and is also mixed-valent (Tl$^{1+}$ and Tl$^{3+}$)
in the parent compound CsTlCl$_3$.
Thus, starting from CsAuCl$_3$, we find a new compound CsTlCl$_3$, which is mixed-valent and has strong phonon coupled bands near the Fermi level,
just like the BaBiO$_3$ compound.

\section{Initial inspection}
To obtain a rough estimation of the lattice constants at ambient pressure, we optimize
the structure of CsTlF$_3$ and CsTlCl$_3$ in the perfect perovskite structure (i.e., without any distortion)
with LDA, GGA and HSE06 functional.
The results are 4.64, 4.82, 4.74 $\AA$ for CsTlF$_3$ and 5.40, 5.61, 5.53 $\AA$ for CsTlCl$_3$, respectively.
It is known that LDA usually underestimates the equilibrium lattice constants while GGA overestimates them.
The fact that the HSE06 optimized values of the lattice constants lie between the LDA and GGA values suggests
the HSE06 functional provides better description of these compounds, similar to the case of BaBiO$_3$\cite{Kresse}.

We next examine the band structures with respects to the F-breathing distortions.
As expected, the REPME of the F breathing mode
in CsTlF$_3$ with lattice constant $a$=4.6 $\AA$ is large as shown in fig.\ref{band}(d),
about 5.0 eV/$\AA$ in DFT-GGA and 8.8 eV/$\AA$ in DFT-HSE06.
The corresponding REPMEs of CsTlF$_3$ and CsTlCl$_3$ at other lattice constants are shown in table \ref{lambda-and-Tc} and \ref{lambda-and-Tc2}.
These large REPMEs indicate a strong electron-phonon coupling of the F breathing mode
which is further enhanced by correlation effects, the same as BaBiO$_3$.
Therefore the CsTlF$_3$ family can be another member of
the ``other high-temperature superconductors''.\cite{BBO}

We further find that there is no information on the CsTlF$_3$ and CsTlCl$_3$ compounds in the ICSD database\cite{ICSD},
thus they are candidates for new materials.
However, the ICSD database\cite{ICSD} has many literature reporting successful synthesis of
the $A_2$Tl$M$F$_6$ compounds ($A$ is an alkaline metal and $M$ is a metal element other than Tl)
which crystallize in the cubic perovskite structure.
This suggests it is possible to synthesize the CsTlF$_3$ and CsTlCl$_3$ compounds and other members in this family.

To check the energetic stability, we consider here a possible reaction to synthesize CsTlF$_3$ as an example:
2CsF+TlF+TlF$_3$$\rightarrow$2CsTlF$_3$+$Q$, where all the compounds on the left hand side,
i.e., CsF, TlF and TlF$_3$, are readily available experimentally and $Q$ is the emitted energy during the reaction.
Our DFT-GGA calculations suggest the reaction releases about 1.1 eV energy per CsTlF$_3$ unit (106 kJ/mol) 
and confirm that the above reaction is energetically favored.
In constrast, we consider also a currently known compound Cs$_3$TlF$_6$ and a possible synthesis reaction:
3CsF+TlF$_3$$\rightarrow$Cs$_3$TlF$_6$+$Q$. This reaction actually absorbs about 1.7 eV energy per Cs$_3$TlF$_6$ unit (164 kJ/mol) according to our
DFT-GGA calculations. Therefore, CsTlF$_3$ is more energetically favored than the existing Cs$_3$TlF$_6$ compound.

\section{Lattice stability and mapping the phase boundary}
Without doping, the CsTlF$_3$ and CsTlCl$_3$ have strong F/Cl breathing distortions which makes the compounds insulating.
To make superconductivity possible, the CsTlF$_3$ and CsTlCl$_3$ have to be properly doped and compressed to suppress the lattice distortions.

To check the lattice dynamics and optimize the doping and pressure, we carried out linear response theory (LRT) calculations
using the LMTART code\cite{lmtart1} with local density approximation functional\cite{PW91}.
The calculations are done using the simple cubic structure with one f.u. per unit cell
at a few hole doping levels and with a few lattice constants (pressure) for both compounds.
The virtual crystal approximation is used to simulate the doping effect.

In the LRT-LDA calculations, we find for CsTlF$_3$ unstable phonon modes 
near the zone boundary at lattice constant $a$=4.6 $\AA$ for all the hole doping
levels studied (0.3, 0.35, and 0.4 hole/f.u.), which indicates that the compound is still in the CDW phase with large hole doping.
Whereas at $a$=4.5 $\AA$, there is no unstable phonon mode in CsTlF$_3$ for the corresponding doping levels.
LDA and the virtual crystal approximation underestimate 
the tendency towards charge disproportionation and structural distortion, thus provide a lower bound 
for the doping at which the material metallizes. 
This methodology also suggests that applying pressure is an efficient way to suppress them to enter the metallic phase.
Using supercell frozen phonon calculations, we further check the dynamic stability of a hypothetical CsXeTl$_2$F$_6$ compound
(a 0.5 hole/f.u. doped CsTlF$_3$) with respect to F breathing distortion. A 2$\times$2$\times$2
supercell is constructed with half of Cs atoms replaced by Xe atoms in order to calculate the phonon frequency of the F breathing mode at $R$ point.
The resulting phonon frequency is 23.5 meV in DFT-GGA and 15.2 meV in DFT-HSE06 with lattice constant $a$=4.5 $\AA$.
Therefore, the F breathing distortion/instability in the parent compound at this volume disappears 
after certain amount of hole doping in both the GGA and HSE06 treatments.
The above calculations suggests that the critical lattice constant to 
induce metallicity and possibly superconductivity lies between 4.5-4.6 $\AA$ for CsTlF$_3$.
Similarly, we find the critical lattice constant for CsTlCl$_3$ is about 5.4 $\AA$.

\begin{figure}[htb]
\includegraphics[width=0.80\linewidth]{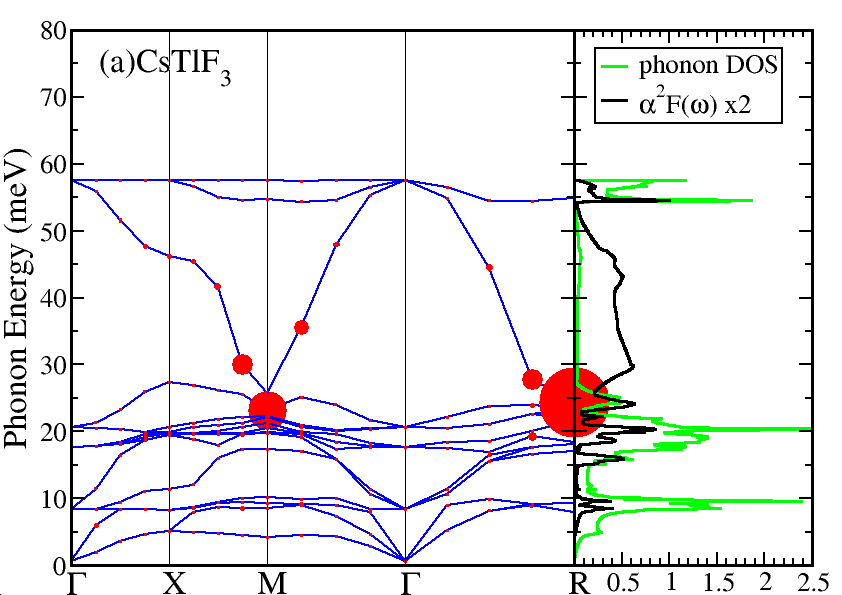}
\includegraphics[width=0.80\linewidth]{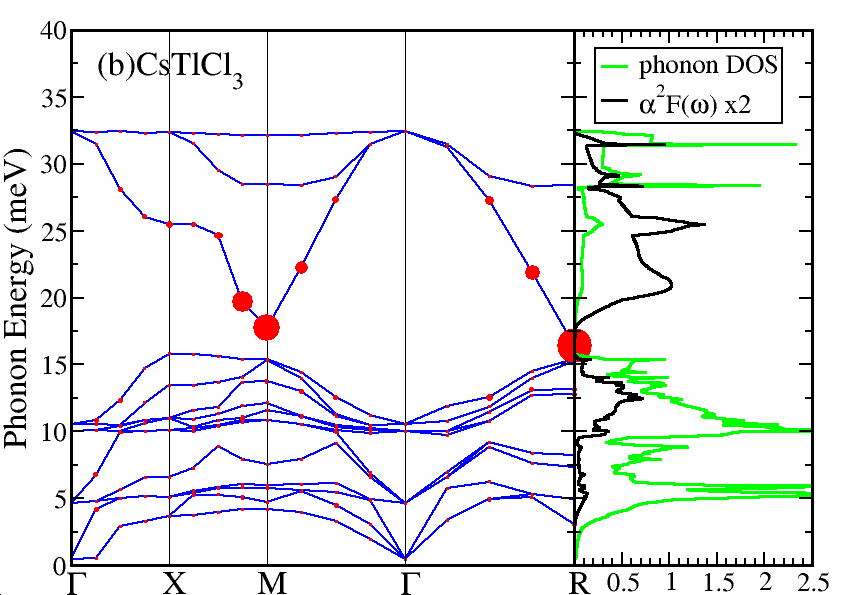}
\caption{
(Color online)
\textbf{Lattice dynamics and electron-phonon coupling.}
The LRT-LDA calculated (left panel) phonon spectra, mode and momentum-dependent
electron-phonon coupling $\lambda$ whose value is proportional to the radius of the corresponding red dot,
(right panel) phonon density of state and $\alpha^2F(\omega)$
of (a) simple cubic CsTlF$_3$ at $a$=4.5 $\AA$ and (b)simple cubic CsTlCl$_3$ at $a$=5.3 $\AA$
both with 0.35 hole doping/f.u. (virtual crystal approximation).
}
\label{phonon1}
\end{figure}

\begin{table*}[htb]
\caption{
\textbf{The calculated results of selected quantities for simple cubic $A$TlF$_3$ ($A$=Cs, Rb, and K).} The displayed items from left to right are
the lattice constant $a$ ($\AA$), the corresponding pressure (GPa) for $A$=Cs, Rb, and K calculated by HSE06 using the simple cubic structure containing 1 f.u.,
the reduced electron-phonon matrix element (REPME, eV/$\AA$) of the F-breathing mode (see fig.\ref{band}a and c) calculated by GGA and HSE06,
the hole doping level $x$ per formula unit in the LRT-LDA calculations where the virtual crystal approximation is used for doping,
the LRT-LDA calculated average phonon frequencies $\omega_{log}$ (K),
the LRT-LDA calculated electron-phonon coupling $\lambda$ contributed by optical phonon modes
and the total electron-phonon coupling $\lambda$, the adjusted total $\lambda$ for the HSE06 approach,
the estimated T$_c$ (K) for LRT-LDA and HSE06 approaches.
The lattice dynamical results are based on the assumption that the real material is stable at the proposed doping/pressure/structure.
The critical temperatures are obtained from
the revised Allen-Dynes formula\cite{Allen} using $\mu^*$=0.10.
The total $\lambda$ and T$_c$ of HSE06
are obtained by rescaling the LRT-LDA calculated $\lambda$ and the average frequencies following ref.~\cite{BBO}.
LRT-LDA and HSE06 are denoted as LDA and HSE, respectively. 
Except for the pressure, other quantities are shown for CsTlF$_3$ only considering that the $A$ elements has little effects on the electron-phonon interaction related properties.
The lattice constants are about 4.74, 4.69, and 4.66 $\AA$ at ambient pressure for $A$=Cs, Rb, and K, respectively, according to HSE06 calculations.
 }
\label{lambda-and-Tc}
\begin{tabular}{|c|c|cc|c|c|ccc|cc|}
\hline
$a$ ($\AA$) & Pressure (GPa)   &\multicolumn{2}{c|}{REPME (eV/$\AA$)} & $x$ & $\omega_{log}$ (K) & \multicolumn{3}{c|}{$\lambda$} & \multicolumn{2}{c|}{T$_c$ (K)} \\
    & HSE (Cs/Rb/K) &GGA & HSE &   &  & LDA opt & LDA tot & HSE & LDA & HSE  \\
\hline
4.60 & 5.4/3.3/2.3 &5.0 & 8.8 &     &       &       &     &    &   &     \\
\hline
4.50 & 12/9.2/7.7  &5.5 & 9.6 & 0.35 & 293  & 0.47           & 0.55         & 1.51          & 5.2      & {\bf 30} \\
\hline
4.40 & 19/16/14 &6.1 & 10.3 & 0.35 & 377    & 0.26           & 0.32         & 0.81          & 0.3      & {\bf 16} \\
\hline
\end{tabular}
\end{table*}

\begin{table*}[htb]
\caption{The same quantities as in table~\ref{lambda-and-Tc} but for the simple cubic $A$TlCl$_3$ compound ($A$=Cs, Rb, and K).
Except for the pressure, other quantities are shown for CsTlCl$_3$ only.
The corresponding lattice constants at ambient pressure are about 5.53, 5.50, 5.48 $\AA$ for $A$=Cs, Rb, and K, respectively.
 }
\label{lambda-and-Tc2}
\begin{tabular}{|c|c|cc|c|c|ccc|cc|}
\hline
$a$ ($\AA$) & Pressure (GPa)   &\multicolumn{2}{c|}{REPME (eV/$\AA$)} & $x$ & $\omega_{log}$ (K) & \multicolumn{3}{c|}{$\lambda$} & \multicolumn{2}{c|}{T$_c$ (K)} \\
    & HSE (Cs/Rb/K) &GGA & HSE &   &  & LDA opt & LDA tot & HSE & LDA & HSE  \\
\hline
5.40  & 2.2/1.6/1.3 & 4.9 & 8.1 & 0.35 & 142 &  0.80           & 0.91         & 2.32          & 9.0      & {\bf 21} \\
\hline
5.30  & 4.5/3.8/3.5 & 5.2 & 8.4 & 0.35  & 191 &  0.47          & 0.53         & 1.30          & 3.1      & {\bf 17} \\
\hline
5.20  & 7.4/6.4/5.8 & 5.5 & 8.8 & 0.35  & 213 &  0.35          & 0.40         & 0.94          & 0.9      & {\bf 12} \\
\hline
5.10 & 11/9.9/9.3   & 5.8 & 9.3 & 0.35  & 222 & 0.26          & 0.33         & 0.73          & 0.3      & {\bf 7.8}  \\
\hline
\end{tabular}
\end{table*}

\section{Lattice dynamics, electron-phonon coupling, and T$_c$}
Now we proceed to report the lattice dynamical properties from the LRT-LDA calculations and
estimate the electron-phonon coupling $\lambda$ and the superconducting T$_c$ of hole doped CsTlF$_3$ and CsTlCl$_3$.
The LRT-LDA method has been shown to be very reliable in predicting T$_c$ for conventional superconductors\cite{deepa, yttrium, calcium, mgb2}
and is a good starting point for exotic superconductors with correlation-enhanced superconductivity. 
The LRT-LDA slightly overestimates the average phonon frequency in the ``other high temperature superconductors" and 
we renormalize it by $\sqrt{(1+\lambda)}$.\cite{BBO} The realistic electron-phonon coupling $\lambda$ is evaluated from 
the LRT-LDA calculations and frozen phonon supercell calculations as described in details in ref.\cite{BBO}. 

In fig.\ref{phonon1}, we show the LRT-LDA calculated phonon spectra, phonon density of states and the Eliashberg function $\alpha^2F(\omega)$
for CsTlF$_3$ ($a=4.50 \AA$) and CsTlCl$_3$ ($a=5.30 \AA$) with 0.35 hole doping/f.u..
In the phonon spectra, we denote each frequency at each $q$ point by a dot whose radius is proportional to the mode- and momentum dependent
electron-phonon coupling $\lambda(q, \omega)$ of the corresponding phonon mode.
All the calculated phonon modes at all $q$ points have positive phonon frequencies
therefore the 0.35/f.u. hole-doped CsTlF$_3$ and CsTlCl$_3$ compounds in the cubic perovskite structure at the studied volumes
are dynamically stable at the LRT-LDA level.
A common feature of the phonon spectra is that the phonon frequency of the F-breathing mode
strongly softens from the zone center $\Gamma$ point to the zone boundary $M$ and $R$ points.
For example, 
the phonon frequency of the F-breathing mode strongly softens from about 60 meV at $\Gamma$ point to
only $\sim$23 meV at $M$ and $R$ points in 0.35/f.u. holed doped CsTlF$_3$ as shown in fig.~\ref{phonon1}(a).
Accompanying the softening of the phonon frequency,
the F-breathing phonon modes acquire large electron-phonon coupling when approaching the $M$ and $R$ points.
This suggests that the F-breathing mode makes a major contribution to the total electron-phonon coupling strength $\lambda$,
which is further supported from the Eliashberg function $\alpha^2F(\omega)$ shown in fig.\ref{phonon1}(a).
The same feature discussed above is found in CsTlCl$_3$ as well,
as shown in fig.~\ref{phonon1}(b), whereas the phonon frequencies in CsTlCl$_3$ are much smaller than
CsTlF$_3$, due to the heavier atomic mass and bigger radius of Cl atom (thus larger equilibrium lattice constant).

In table \ref{lambda-and-Tc} and \ref{lambda-and-Tc2},
we present, respectively, the calculated results of some important quantities
for CsTlF$_3$ and CsTlCl$_3$ at various lattice constants and doping levels.
For each lattice constant, the corresponding pressure is calculated by DFT-HSE06 
with undistorted perovskite structure of the undoped compound.
The other quantities included are the REPMEs of the F/Cl breathing mode using DFT-GGA and DFT-HSE06,
the average phonon frequency $\omega_{log}$,
the electron-phonon coupling $\lambda_{L,o}$ (contributed by optical phonons only)
and $\lambda_L$ (contributed by all phonons) from the LRT-LDA calculations,
the estimated total $\lambda$ within DFT-HSE06,
and the superconducting temperature T$_c$.
Since the electron-phonon coupling can be very strong ($\lambda>1$),
we adopt the revised Allen-Dynes formula\cite{Allen} with $\mu^*$=0.1 to estimate T$_c$.

We find in the LRT-LDA calculations, that at a fixed volume, 
the total electron-phonon coupling $\lambda$ (average frequencies) generally increases (decrease)
with decreasing doping level, in accordance with the trend shown in the insert of fig.~\ref{phase-diagram}
and consistent with the expectation that the simple cubic perovskite structure become unstable
towards smaller doping levels.
At a fixed doping level, the total electron-phonon coupling $\lambda$ 
generally decreases with decreasing volume (increasing pressure)
in both CsTlF$_3$ and CsTlCl$_3$ whereas the average phonon frequencies show the opposite trend,
namely, they increase with decreasing volume (increasing pressure) as expected.
Therefore, the maximum T$_c$ with varying volume (pressure) relies on the competition between $\lambda$ and $\omega_{log}$.
For the volumes we studied, the LRT-LDA theoretical electron-phonon coupling is as high as 1 
and the maximum T$_c$ reaches about 10 K for CsTlF$_3$ and CsTlCl$_3$, which already suggest
that these compounds are good phonon-mediated superconductors.
By taking into account non-local correlation effects,
the maximum theoretical T$_c$s estimated by HSE06 are 30 K and 21 K
for optimal doped CsTlF$_3$ and CsTlCl$_3$ compounds under about 12 and 2 GPa pressure, respectively.
(See table.~\ref{lambda-and-Tc} and \ref{lambda-and-Tc2})

\section{Discussion and Conclusion}
Our results support the validity of the qualitative phase
diagram shown in fig.\ref{phase-diagram}, and provide
quantitative estimates, which should motivate the experimental
synthesis of the materials described in this letter.
In this class of materials it is important to optimize the volume/pressure
to achieve maximum T$_c$. 
At large volumes/low pressures, the CDW
phase is so stable that it cannot be suppressed by reasonable
doping. On the other hand, at very small volumes/high pressures,
the electron-phonon coupling become weak and the T$_c$ is
vanishingly small. The critical superconducting temperature T$_c$ as
high as 30 K and 20 K are expected in 
optimally doped and compressed 
CsTlF$_3$ and CsTlCl$_3$, respectively,
within the uncertainty of the calculations, 
which are meant to provide guidance to the experimental search for new superconductors 
and can be refined by performing larger supercell calculations.

In addition to applying pressure, the volume of the CsTlF$_3$ and
CsTlCl$_3$ compounds can be tuned by replacing the Cs atoms with
other alkaline atoms such as Rb and K. The HSE06 optimized
lattice constants at zero pressure for RbTlF$_3$, KTlF$_3$,
RbTlCl$_3$, and KTlCl$_3$ are about 4.69, 4.66, 5.50, 5.48 $\AA$,
respectively. On the other hand, replacing F and Cl with Br and I
is detrimental for superconductivity due to their heavier masses.

To reach a doping of 0.3-0.4 hole/f.u., it is best to substitute
alkaline atoms with vacancies and/or charge neutral units. Doping
on the active sites by replacing F/Cl atoms with O/S and/or N/P
atoms or substituting Tl by Hg, introduces disorder and removes
the active pairing sites. Other avenues for doping should also be explored
such as electrostatic gating which was shown recently in ref.\cite{Ueno} 
to be able to induce superconductivity in KTaO$_3$ at 50 mK.

Our approach for designing 
novel high-temperature superconductors is quite general and can
be used to search for other ``other high temperature
superconductors".

\acknowledgments Z.P.Y and G.K. were supported by the AFOSR-MURI
program towards better and higher temperature superconductors. We
are grateful to Mac Beasley and Martha Greenblatt 
for discussions and a critical reading
of the manuscript.

\end{document}